%% file: main.tex
\documentclass[conference]{IEEEtran}
\IEEEoverridecommandlockouts
\usepackage{cite}
\usepackage{amsmath,amssymb,amsfonts}
\usepackage{algorithmic}
\usepackage{graphicx}
\usepackage{textcomp}
\usepackage{xcolor}
\usepackage{booktabs}
\usepackage{xspace}
\usepackage{amssymb,amsmath}
\usepackage{balance}
\usepackage{flushend}
\usepackage{subcaption}
\usepackage{tcolorbox}
\usepackage{multicol}
\usepackage{multirow}
\usepackage[export]{adjustbox}
\usepackage{hyperref}
\usepackage{verbatim}
\usepackage{float}
\usepackage{adjustbox}
\usepackage{longtable}

\usepackage{listings}
\lstdefinelanguage{diff}{
  language=Java,
  morecomment=[f][\color{blue}]{@@},     
  morecomment=[f][\color{red!60!black}]-,         
  morecomment=[f][\color{green!60!black}]+,       
  morecomment=[f][\color{magenta}]{---}, 
  morecomment=[f][\color{magenta}]{+++},
}

\newcommand{\toolname}{\text{$\mu$\textsc{IntMut}}\xspace}
\newcommand{\llmo}{\text{$o$\textsc{IntMut}}\xspace}

\newcommand{\mbert}{\text{$\mu$\textsc{Bert}}\xspace}

\def\BibTeX{{\rm B\kern-.05em{\sc i\kern-.025em b}\kern-.08em
    T\kern-.1667em\lower.7ex\hbox{E}\kern-.125emX}}

\begin{document}

\title{	Intent-Based Mutation Testing: From Naturally Written Programming Intents to Mutants}

\author{
\IEEEauthorblockN{Asma Hamidi}
\IEEEauthorblockA{\textit{SnT, University of Luxembourg} \\
Luxembourg \\
asma.hamidi@uni.lu}
\and
\IEEEauthorblockN{Ahmed Khanfir}
\IEEEauthorblockA{\textit{Medtech, South Mediterranean University} \\
Tunis, Tunisia \\
ahmed.khanfir@medtech.tn}
\and
\IEEEauthorblockN{Mike Papadakis}
\IEEEauthorblockA{\textit{SnT, University of Luxembourg} \\
Luxembourg \\
michail.papadakis@uni.lu}
}

\maketitle

\begin{abstract}
This paper presents intent-based mutation testing, a testing approach that generates mutations by changing the programming intents that are implemented in the programs under test. In contrast to traditional mutation testing, which changes (mutates) the way programs are written, intent mutation changes (mutates) the behavior of the programs by producing mutations that implement (slightly) different intents than those implemented in the original program. The mutations of the programming intents represent possible corner cases and misunderstandings of the program behavior, i.e., program specifications, and thus can capture different classes of faults than traditional (syntax-based) mutation. Moreover, since programming intents can be implemented in different ways, intent-based mutation testing can generate diverse and complex mutations that are close to the original programming intents (specifications) and thus direct testing towards the intent variants of the program behavior/specifications. We implement intent-based mutation testing using Large Language Models (LLMs) that mutate programming intents and transform them into mutants. We evaluate intent-based mutation on 29 programs and show that it generates mutations that are syntactically complex, semantically diverse, and quite different (semantically) from the traditional ones. We also show that 55\% of the intent-based mutations are not subsumed by traditional mutations. Overall, our analysis shows that intent-based mutation testing can be a powerful complement to traditional (syntax-based) mutation testing.
\end{abstract}


\section{Introduction}
Mutation testing has long been recognized as one of the most powerful testing techniques \cite{ChekamPTH17, PapadakisK00TH19}. It generates program variants by altering the way programs are written, i.e., by making simple syntactic changes to the code under test. These variants are then used as targets for differential program analysis, that is, test writing (or test selection) with the aim to distinguish the behavior of the original program from that of the variants. When a test triggers a difference in the behavior of the mutant and the original programs, the mutant is considered as covered, called 'killed', otherwise is considered as not covered and called 'live'. The effectiveness of the test suites is then measured by the mutation score, the proportion of mutants killed over all considered mutants \cite{PapadakisK00TH19}.
 
Traditional mutation testing operates at the program syntax level and thus is typically oriented toward errors that are syntactically small, i.e., the syntactic distance of the variants to the original program is rather small. For example, a typical mutation is to replace one operator such as '$>$' with another '$>=$' in a relational expression. This approach allows the introduction of subtle semantic deviations that make mutation effective at testing the behavioral boundaries of the programs under test. At the same time, this approach limits testing to the program logic that is actually implemented, making mutation testing less effective in revealing complex behavior-oriented (falling on the core of business logic) and omission faults \cite{ChekamPTH17}. 

To address these issues, we propose a novel approach to mutation testing, namely \textit{intent-based mutation testing}. An intent is the programmer's objective for the code, described informally in natural language and offers a description of the task that is implemented. For example, if a programmer intends to create a function that calculates the factorial of a number, the intent could be formulated as  "a function that takes an integer as an input and calculates its factorial". Intent-based mutation aims at testing programs by formulating alternative implementations of the same intents (intents implemented in the original program) as well as by formulating intent variant implementations corresponding to slightly altered intents (mutated intents). In other words, we generate mutated intents and contrast their respective implementations. We consider that intent-based mutation testing can target potential misunderstandings of the program's intents or specifications, which include faults that are hard to capture with traditional mutation.

By making minor adjustments to the initial intents of the program under test, it is possible to introduce mutations that reflect (small) misunderstandings of the actually implemented programming intents. These adjustments lead to an implementation that is being interpreted differently from what was originally done. Additionally, mutated intents can lead to mutations that include global transformations spanning across the entire intent implementation, e.g., an entire method. This approach contrasts with traditional mutation testing, where small and local changes are made to the programs syntax. 

Intent-based mutation testing aims to find a different class of faults and should complement traditional mutation testing. When dealing with a programming intent, intention-based mutation can be seen as a process that mutates the intended behavior/specifications rather than the program. Previous work has investigated ways to create mutations considering the code context of mutation points \cite{ChekamPBTS20,tcap, DeepMutation}, but these approaches are fundamentally limited to traditional syntactic mutations and therefore share the limitations of traditional mutation testing. 

We release intent-based mutation testing by using automatic programming tools, such as Large Language Models, to automatically formulate program implementations based on programming intents written in natural language. We generate intent variants by asking the tools to directly generate variant implementations and by mutating the natural language descriptions, which are then turned into actual programs and form our intent-based mutations.

 We evaluate intent-based mutation on 29 HumanEval+ programs and show that it generates syntactically complex mutations that are semantically different from traditional mutations and diverse. Perhaps more importantly, we also show that more than 23\% of the intent-based subsuming mutations cannot be detected by any of the traditional mutation-killing tests. Overall, our analysis corroborates the finding that intent-based mutations are strong and introduce faults that are not captured by traditional mutation testing techniques.
 
This paper makes the following contributions:
\begin{itemize}
  \item We present intent-based mutation testing, outline its role in testing, discuss its difference from traditional mutation testing, and detail how it can be implemented. We also outline future research directions towards semantic and specification-driven testing approaches. 
  \item We show the ability of intent-based mutation testing to generate complex and semantically diverse mutations, which cannot be detected by traditional mutation-based tests, i.e., tests generated to kill traditional mutations.
  \item We show that intent-based mutation has the potential to expand the fault detection abilities of mutation testing by revealing 23\% more faults than those revealed by traditional mutation.
\end{itemize}
    
\section{Background and related work}

\subsection{Mutation Testing}

Mutation testing typically operates by introducing a few changes to the program code, thereby creating many different versions of it (named \emph{mutants})~\cite{PapadakisK00TH19}. 
Those changes are usually obtained by applying predefined patterns called \emph{mutation operators}~\cite{PapadakisK00TH19} on the target code, which can, for example, invert relational operators (e.g., replacing $\geq$ with $<$) or arithmetic operators (e.g., replacing $+$ with a $-$), etc.
Mutants can be used to assess the strength of test suites, by measuring their ability to trigger different behaviors from the original program. 
If a test suite fails when executed on a mutant, it is said to be \textit{killed}, else, it is said to be \textit{live} or \textit{survived}.  
As some mutants cannot be killed, i.e., if they are functionally equivalent to the original program, they are said to be \textit{equivalent}, otherwise, they are said to be \textit{killable} \cite{MarcozziBKPPC18}. 
By computing the ratio of killed mutants by a test suite (among all the generated ones), we can measure the test suite adequacy. This ratio is called the \emph{mutation score}.
The live mutants can serve as testing objectives and guide developers in writing effective tests \cite{MarcozziBKPPC18}. Mutation testing techniques generate redundant mutants that can be \textit{duplicated} \cite{PapadakisJHT15} or \textit{equivalent} \cite{MarcozziBKPPC18, KintisPM15} to the original program. 
 
Much research have focused on improving the efficiency of mutation testing, aiming at increasing the coverage, diversity and real-fault coupling while reducing the redundancy among the generated mutants. This has resulted in the proposal of several pattern-based mutation techniques~\cite{mujava,major,pitest} that rely on syntactic mutation operators. These operators have been designed mainly on the basis of the target programming language grammar and have been empirically tuned through multiple studies~\cite{0020331, PapadakisK00TH19,OffuttLRUZ96, MarcozziBKPPC18, KintisPPVMT18} to increase their effectiveness. 

With the advancement of machine learning, recent research has focused on generating mutants based on real faults. For instance, Tufano et al.~\cite{DeepMutation} and Zao et al.~\cite{tian2022learning} proposed neural machine translation techniques to inject faults, trained on real bug fixes. 
Patra et al.~\cite{SemSeed} proposed also a learning approach, that adapts then applies pre-learned fault patterns on the target project. Khanfir et al.~\cite{khanfir2020ibir} proposed the usage of bug reports together with inverted automated-program-repair operators to inject faults. 
Their results are promising, however, may be of limited usability, depending on the availability of good bug reports or diverse and untangled fix commits~\cite{herzig2011untangling}. 

Degiovanni et al. ~\cite{DBLP:conf/icst/DegiovanniP22,khanfir2023efficient} proposed \mbert a context-aware mutation testing technique which does not rely on historical bugs or the language grammar but rather on LLM, i.e. CodeBERT~\cite{DBLP:conf/emnlp/FengGTDFGS0LJZ20}, knowledge of developer code. This approach mutates the target program by replacing its tokens one at once with inaccurate CodeBERT predictions, producing several likely-to-occur mutants. 
Empirical comparative studies~\cite{ojdanic2023syntactic,ojdanic2023comparing} with other learning and pattern-based approaches, give evidence of its high efficiency and cost efficiency in generating mutants that couple and reveal real faults, which makes it a suitable comparison baseline for our approach.  

Unlike those approaches, our approach does not apply changes to the program code, but to its intent, written in natural language, instead. Hence, it does not depend on any prior particular knowledge, i.e. historical real bugs or programming language grammar. In fact, it relies solely on the natural language comprehension and code generation capability of LLMs.


\begin{figure*}[t]
    \centering
    \includegraphics[width=0.92\textwidth]{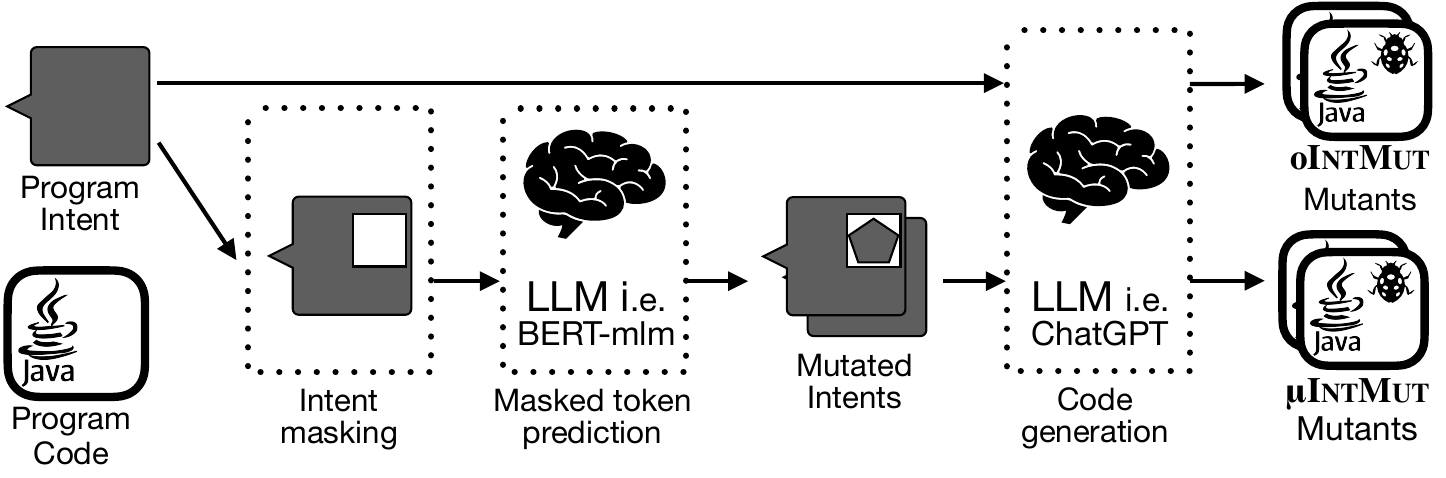}
    \caption{Intent-based mutation testing workflow.}
    \label{fig:pipeline}
\end{figure*}
\subsection{Large Language Models}

Large Language Models (LLMs)~\cite{copilot,openAIcodex,codewhisperer}, such as GPT (Generative Pre-trained Transformer), have revolutionized the field of natural language processing (NLP) with their ability to understand and generate human-like text. One of the remarkable capabilities of these models is their efficiency in generating code from natural language descriptions. GPT models, in particular, can interpret a user's intent expressed in plain language and translate it into executable code across various programming languages. This is achieved through extensive training on diverse datasets, including code repositories and documentation, enabling the model to learn syntax, semantics, and common programming patterns. As a result, developers can leverage GPT to automate coding tasks. 

\section{Approach}

Our approach uses 
BERT~\cite{devlin2018bert} to mutate the natural intent, by replacing its tokens with the inaccurate predictions of the model. We invoke its Masked Language Modeling (MLM) pipeline to predict replacements of masked tokens from the input intent, based on the context of the remaining intent text.
As the model has been trained on a large corpus, and is able to write human-like text, even when inaccurate, we expect its predictions to result into meaningful mutated intents.

We leverage the capability of Large Language Models (LLMs) to link natural language descriptions (task descriptions) and code to generate mutations based on code intents. The mutations are the result of the mutated intents and/or LLM mistakes in generating code for a given intent. A high-level overview of the functioning of our approach is described in Figure~\ref{fig:pipeline}.

This approach operates in 4 distinct steps as follows:
\begin{enumerate}
    \item It masks tokens from the given input intent, creating several masked versions of it -- one for each token.
    \item Then, it passes those intents to a language model, i.e., BERT, to predict a value for each masked token. 
    \item Next, it invokes an LLM, i.e. GPT, to generate an implementation for each of those generated intents.
    \item In addition, our approach generates mutations, by generating code directly (step 3) using the original intent but asking for multiple, i.e., 10, alternative implementations.
\end{enumerate}

\subsection{Intent masking}
In this step, we mask the intent description tokens, one at a time, producing one masked intent per token. This means that every masked version contains the original intent with one missing token, replaced by the placeholder~\texttt{<mask>}. 

This way, we can obtain several mutations from the same intent with small syntactical differences (one token difference). As we aim at introducing behavioural mutations, we exclude tokens that are irrelevant to the text context, i.e. the punctuation characters, and mask only alphanumeric tokens.

For example, for the sentence \textit{function that takes an integer}, {\toolname} produces the following masked sequences: 
\begin{itemize}
    \item \textbf{\texttt{<mask>}} that takes an integer

    \item function \textbf{\texttt{<mask>}} takes an integer
    \item function that \textbf{\texttt{<mask>}} an integer
        \item function that takes \textbf{\texttt{<mask>}} integer

    \item function that takes an \textbf{\texttt{<mask>}} 
\end{itemize}

\subsection{BERT-MLM prediction} 
\toolname invokes BERT~\cite{devlin2018bert}, a pre-trained language model, to predict replacements for the masked tokens. 
To do so, it tokenizes every masked version into a tokens vector then crops it to a subset one that fits the maximum size allowed by the model (512). 
Next, our approach feeds these vectors to BERT-MLM to predict the most probable replacements of the masked token.
Our intuition is that the larger the text portion accompanying the mask placeholder, the better BERT would be able to capture the text context, and consequently, the more meaningful its predictions would be. 
This step ends with the generation of one mutated intent per masked token. 

\subsection{LLM code generation}
We employ GPT-3.5- with 0.8 temperature, to generate code from intents written in natural language. We run it with the intents from the previous step and generate a code implementation per mutated intent. As those codes have been generated via different intents, we expect them to be different, and thus useful as mutants. 

The proposed approach produces also mutants by direct invocation of the LLM with the original code intent. 
This means that for a given code intent, it asks the LLM to generate different implementations of it, i.e. ten alternative implementations for a given intent. 

This way we can generate mutants that represent mistakes of the intent done by the LLM. We thus, have two approaches:
\subsubsection{Mutated intents (\toolname{})} Generates mutants by mutating the original intent and then generating their corresponding implementations.
\subsubsection{Original intents (\llmo{})} Generates mutants by generating alternative implementations of the original intent.

\section{Experimental study}




\subsection{Research questions}

Intent-based mutation testing alters the original programming intents with the intention of producing similar but different implementations. To this end, we investigate the extent to which this is possible when using our framework. We thus check whether the intents lead to mutations that are syntactically and semantically different from the original program. An additional aspect we consider is the differences among the mutations themselves, as it is not useful to produce variants that are not diverse. Therefore, we ask:

\textbf{RQ1 (Diversity)}: \emph{Does intent-based mutation testing generate valid mutations that are syntactically and semantically diverse?}

We answer this question by measuring the syntactic and semantic similarity of the intent-based mutations with the original programs. We also check the subsumption relationships among these mutations to study the semantic overlaps between those mutations, as well as the number of mutations that are subsuming, i.e., a typical mutation testing diversity metric \cite{ojdanic2023syntactic, PapadakisCT18}. 

Having answered this question and showed that the intent mutations are actually valid and diverse, we contrast them with the mutations generated by other (syntactic-based) approaches to check whether they actually reflect different types of faults than the syntactic methods. Hence, we ask:

\textbf{RQ2 (Overlap with syntactic mutations)}: \emph{Are intent-based mutations semantically different from those mutants generated by syntactic-based mutations?}

To answer this question, we compare the semantic overlaps between the intent-based mutations with those produced by syntactic-based methods. We further strengthen our comparison; we also investigate the extent to which intent-based mutations subsume (and are subsumed by) syntactic mutations. 

The above investigations aim at checking whether intent-based mutations are actually different from the syntactic ones, which leaves out the question of which approach is most effective and to what extent. Thus, we ask: 

\textbf{RQ3 (Effectiveness)}: \emph{How effective is intent-based mutation testing in comparison to syntactic-based one?}

To answer this question, we form mutation-based test suites, with respect to the compared techniques, and check their ability to kill a reference set of subsuming mutants, i.e., the subsuming mutants of all the compared techniques together.  


\subsection{Experimental setup}


\textit{HumanEval}\cite{chen2021evaluating} is a benchmark created to evaluate the models' ability to generate code. It contains 164 python programming problems written by humans, each paired with a solution and test cases. \textit{HumanEval-x}\cite{zheng2023codegeex} is an extension of HumanEval that contains the same 164 problems but includes four additional programming languages: C++, Java, JavaScript, and Go.  \textit{HumanEval+}\cite{liu2024your} is another extension of HumenEval that uses mutation testing to augment the programs' test suites.

We use the Java entries from HumanEval-x, specifically, we use the problem descriptions in natural language (English) reflecting the intentions to generate the mutants following the intent-based mutation we introduce above. To ensure the thoroughness of our analysis, we augment the Java entries using test cases from HumanEval+. We extract test data and expected results from the Python test code and translate them to Java by adapting data types and dropping incompatible test data. We construct Java test cases, compile, and execute them against the ground truth solution provided by HumanEval-x, keeping only the problems for which the entire test suite pass. 

Unfortunately, our test subjects are small and result in very few mutations for many cases. We thus, need to ensure a reasonable number of mutations for each problem we use, and set a minimum threshold of five killable mutants. This means that we select from the dataset the problems for which each approach produces at least five killable mutants, resulting in a total of 29 problems. Table \ref{tab:summary_statistics} records the summary statistics of the number of tests and the length of the description in terms of characters number. It includes the average, median, maximum, and minimum values for each.

\begin{table}[h]
\centering
\renewcommand{\arraystretch}{1.1} 
\setlength{\tabcolsep}{8pt} 
\caption{\centering Descriptive statistics of the problems we consider}
\label{tab:summary_statistics}
\begin{tabular}{|c|c|c|c|c|}
\hline
Statistic & Mean & Median & Max & Min \\
\hline
Number of Tests & 659.2 & 865 & 1025 & 69 \\
\hline
Length of Description & 629.55 & 547 & 1462 & 291 \\
\hline
\end{tabular}
\end{table}


\subsection{Experimental procedure}






To address \textbf{RQ1}, we generate mutants using \toolname on the dataset problems. We start by mutating the problem description by tokenizing it, masking one token at a time, and using BERT to predict a replacement for it. We prompt GPT-3.5-turbo using each of the mutated descriptions. Additionally, we generate mutants by directly prompting GPT-3.5-turbo to produce different implementations based on the same description. Finally, we mutate the original programs in the dataset using \mbert. 

The goal is to investigate whether the obtained implementations translate into mutants. To achieve this, we study the syntactic validity and examine how syntactically different they are from the original program by computing the syntactic distance between them. This step includes the computation of several metrics: BLEU score, the number of distinct tokens, cosine similarity, and Jaccard similarity to compute the distance, which is the difference 1 - similarity. 

Furthermore, we address semantic diversity based on test execution and subsumption relations between mutants. We run the mutants obtained from \toolname against the test cases and identify the minimal set of mutants that subsume all the others. To quantify this, we calculate the percentage of subsuming mutants relative to the total number of killed mutants. A higher percentage reflects broader differences in mutants behavior, suggesting a more diverse set. 

To answer \textbf{RQ2}, we begin by comparing the different approaches semantically using the results of test failure (assertions violation) by identifying the mutants killed by identical sets of tests and illustrate the overlap. We then merge all the killable mutants generated from the three approaches into one large set of mutants. From this set, we identify the minimal subset of subsuming mutants and compute the contribution of each approach to this set. This analysis shows the relative importance of the mutants of one approach against the others. 

To answer \textbf{RQ3} and determine whether \toolname is more effective than the remaining approaches, we gather all the mutants from all approaches and identify their subsuming subset, which is then used to conduct the comparison. The process entails randomly selecting the minimal test sets that kill all mutants generated by \toolname, \llmo and \mbert one at a time, then using these test sets to kill the mutants of the merged subsuming subset. Since the tests are selected randomly, we repeat this process 100 times to reduce the randomness impact on our results.

\textit{Test selection algorithm:} we start with an empty set and each time, we add a randomly chosen test and check if the mutation score increases (more mutants are killed). If not, we drop the test as it is redundant with respect to the already selected tests. We continue this process until all killable mutants from an approach are killed.

For comparison, we use an objective score (subsuming mutation score), which corresponds to the number of subsuming mutants in the merged set killed by the tests selected according to one approach, divided by the total number of subsuming mutants, and we compare the three approaches based on it.

\section{Results}

\subsection{RQ1: Does intent-based mutation testing generate valid mutations that are syntactically and semantically diverse?}
\noindent
\textbf{Syntactic validity:} The mutant generation results reported in Table~\ref{tab:killed_mutants} show that the three compared approaches produce valid (compilable) mutants.
In fact, the majority (over 80\%) of the implementations obtained by intent-based mutation are valid (91\% by \toolname and 81\% by \llmo), which is about three times the generation validity ratio of \mbert (27.6\%). 
This relatively high ratio encourages the approach's design to rely on instruction-based code generation LLMs, i.e. GPT-3.5-turbo, to generate mutants as most of the generation effort results in syntactically valid code. 

Moreover, from the last column (total killed) of Table~\ref{tab:killed_mutants}, we can see that a large ratio of the generated implementations by the proposed approach behave differently from the original code (failing tests that pass on the original code). This confirms that the proposed approach can generate programs that behave differently from the original one and are thereby useful for mutation testing. 


\begin{table}[!h]
\centering
\renewcommand{\arraystretch}{1.2} 
\setlength{\tabcolsep}{4pt} 
\caption{\centering Number of generated, valid, killed and alive mutants for each approach}
\label{tab:killed_mutants}
\begin{tabular}{|c|c|c|c|c|}
\hline
 & Total Generated & Total Valid & Total Alive & Total Killed \\
\hline
\toolname & 2357 & 2144 & 620 & 1524 \\
\hline
\llmo & 290 & 235  & 24  & 211 \\
\hline
\mbert     & 3608 & 996 &  117 & 879 \\
\hline
\end{tabular}
\end{table}

\noindent
\textbf{Syntactic distance:} Table \ref{tab:syntactic_distance} records the mean syntactic distance of mutants produced by the approaches from the original code. We observe that the intent-based ones are significantly more distant than those obtained by \mbert. For instance, intent-based mutants are about 0.7 bleu distant from the original code, which is over 23 times the bleu distance of \mbert mutants that is 0.031. The same difference is observed when comparing the cosine and jaccard distances; 0.003 and 0.021 for \mbert and 0.2 and 0.3 for intent-based mutants.
This highlights the fact that intent-based mutants are more complex than those produced by \mbert, introducing several changes to the original code. 


\begin{table}[h]
\centering
\renewcommand{\arraystretch}{1.1} 
\setlength{\tabcolsep}{3pt} 

\caption{\centering Mean syntactic distance between the mutated and original code}

\label{tab:syntactic_distance}
\begin{tabular}{|c|c|c|c|c|c|}
\hline
Metric & 1 - BLEU  & Tokens Diff. & 1 - cosine & 1 - jaccard \\
\hline
\toolname & 0.672 & 68.91  & 0.185 & 0.284 \\
\hline
\llmo & 0.724 & 63 & 0.218 & 0.274 \\
\hline
\mbert & 0.031 &  1.296 & 0.003 & 0.021 \\
\hline
\end{tabular}
\end{table}


\noindent
\textbf{Semantic Distance:} Table~\ref{tab:semantic_distance} lists the average semantic distances between the mutants and the original code. This metric reflects the number of tests each mutant fails over the total number of tests. 

The table shows that all approaches produce mutants that are semantically distant from the original code. From the first columns, we can see that \textit{\mbert scores the highest average distance of 0.336 compared to the 0.094 and 0.153 scored by respectively \toolname and \llmo.}
This can be explained by the fact that \mbert produces a higher ratio of killable/valid mutants, as illustrated in Table~\ref{tab:killed_mutants}). 
In fact, when considering only the killed mutants, we see that all approaches have relatively similar average distance from the original program, 0.4 and 0.46 for those generated by \mbert intent-based mutations. This validates our approach ability to generate mutants; code implementations that are behaving differently (semantically different) from the original code, and thus can serve for mutation testing.

So far, our results show that our approach is capable of producing syntactically valid, complex, and semantically different mutations from the original code. 
In the following part, we investigate how diverse these mutants are, that is, how different (semantically) each mutant is from the others.





\begin{table}[h]
\centering
\renewcommand{\arraystretch}{1.1} 
\setlength{\tabcolsep}{3pt} 

\caption{\centering Mean semantic distance between the mutations of each approach and the original program}
\label{tab:semantic_distance}
\begin{tabular}{|c|c|c|c|c|c|}
\hline
Metric & Valid Mutants  & Killable Mutants \\
\hline
\toolname & 0.153 & 0.467  \\
\hline
\llmo & 0.094 & 0.46  \\
\hline
\mbert & 0.336 &  0.4 \\
\hline
\end{tabular}
\end{table}

\begin{figure}[ht]
\centering
\includegraphics[width=0.45\textwidth]{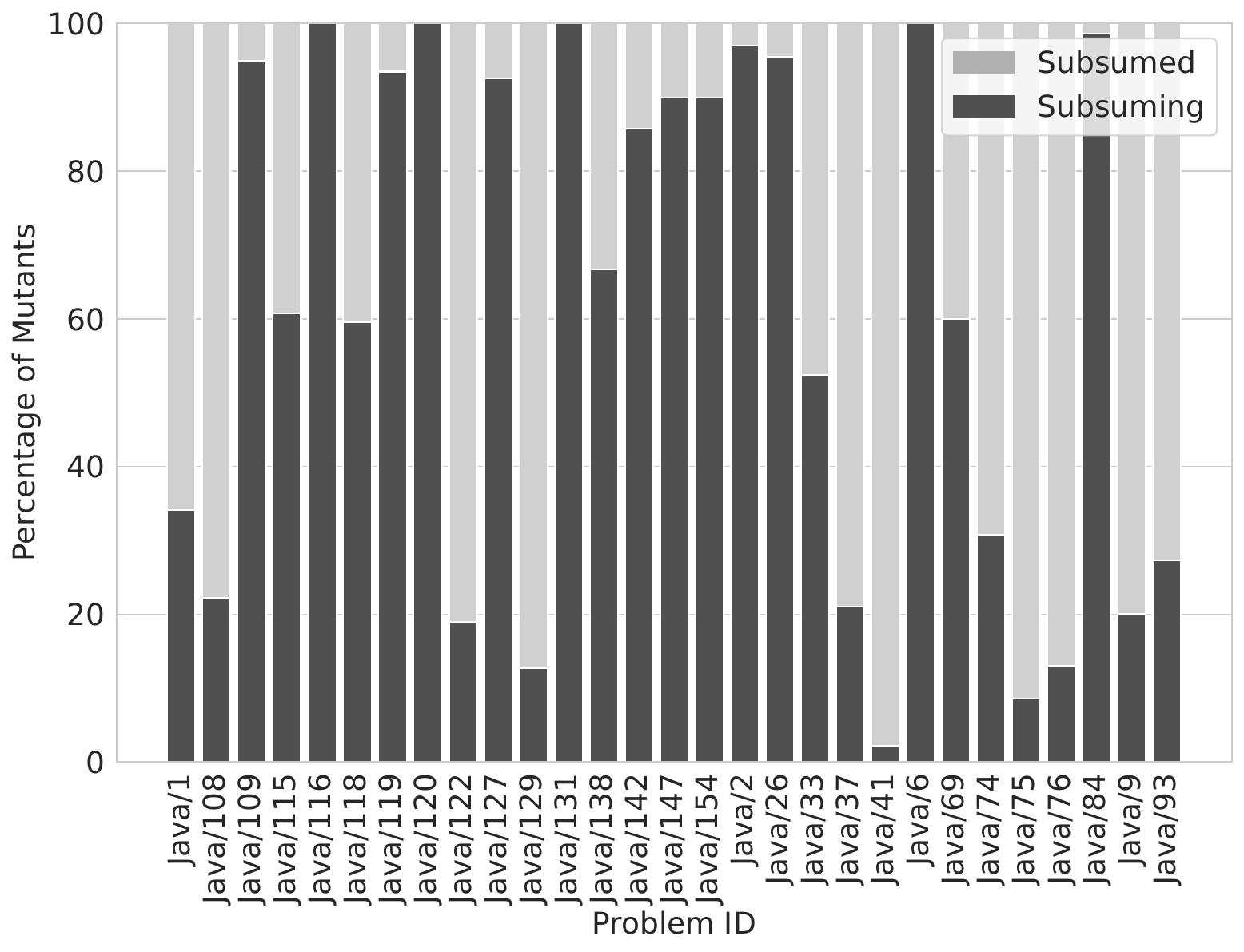}
\caption{\centering Percentages of subsuming (dark grey) and subsumed (light grey) mutants produced by \toolname}
\label{fig:semantic_diversity}
\end{figure}

\noindent
\textbf{Semantic diversity:} We conduct a subsumption analysis \cite{KurtzADOD2014} among the mutations generated by our approach. The plot of Figure~\ref{fig:semantic_diversity} shows the percentage of subsuming and subsumed mutants generated by \toolname. We find that the subsuming mutants form on average 60.2\% and 76.29\% of the mutants killed generated by \toolname and \llmo, respectively.
These ratios are relatively high compared to the subsumed mutants of \mbert that are 42.54\%. This endorses the diversity of the generated mutants by the proposed approach.

When Computing the minimal subsuming mutants set by removing those that break the same test sets, the average proportion of minimal subsuming mutants is 11\% and 24.74\% of killed mutants generated by \toolname and \llmo respectively. This indicates considerable ratio of subsuming mutants being mutually subsumed, with an average of 71\% and 60.58\% of all subsuming mutants for \toolname and \llmo respectively.


\subsection{RQ2: Are intent-based mutations semantically different from the syntactic-based ones?} 

\begin{figure}[!h]
\centering
\adjincludegraphics[width=0.37\textwidth, trim={{.08\width} {.10\width} {.02\width} {.02\width}} ,clip]
{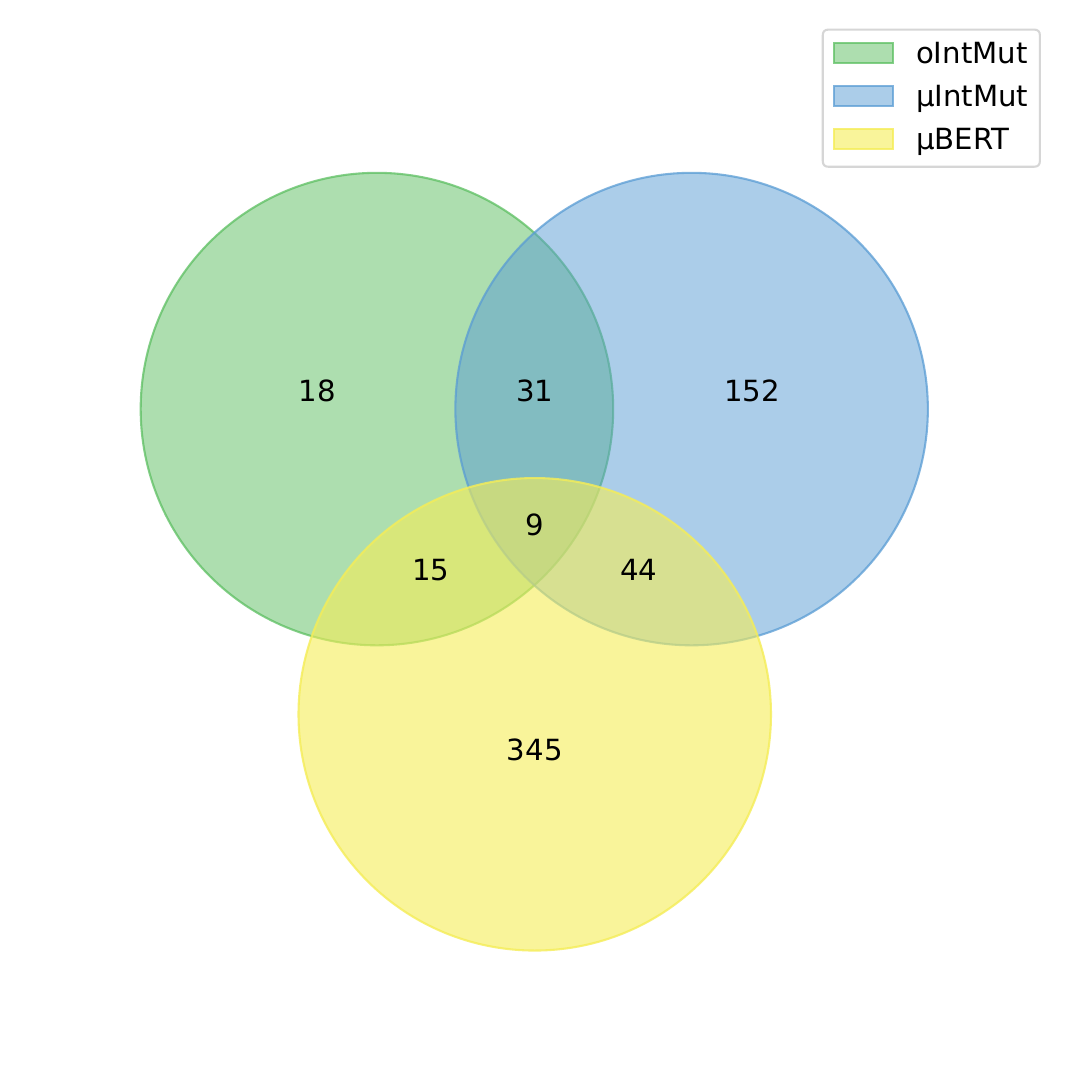}
\caption{\centering Semantic overlap between the studied approaches. The majority of Intent-based mutations are not covered by traditional techniques, showing that it produces different types of faults compared to syntax-based mutation approaches.}
\label{fig:semantic_overlap}
\end{figure}

\noindent
\textbf{Semantic Overlap:} To study the diversity and overlap of mutants generated by each approach, we consider the differences of the killing tests between the mutant we study. This means that a mutation is unique if it is killed by a test set that is not killing any other mutant. While mutants that are killed by the same set of tests are considered an overlap. To illustrate the overlap, we begin by removing all the semantic duplicates in each approach before computing the overlap.
Figure~\ref{fig:semantic_overlap} illustrates a Venn diagram of the unique mutants generated by each approach.


From the diagram, we observe that \llmo has the lowest rate of semantically unique mutants, with 24.6\% of its killable mutants being detected by a unique set of tests, compared to 64.4\% and 83.3\% of unique mutants in \toolname and \mbert respectively. This shows that intent-based mutation produces different types of faults compared to syntax-based mutation.

\begin{figure}[!h]
\centering
\includegraphics[width=0.45\textwidth]{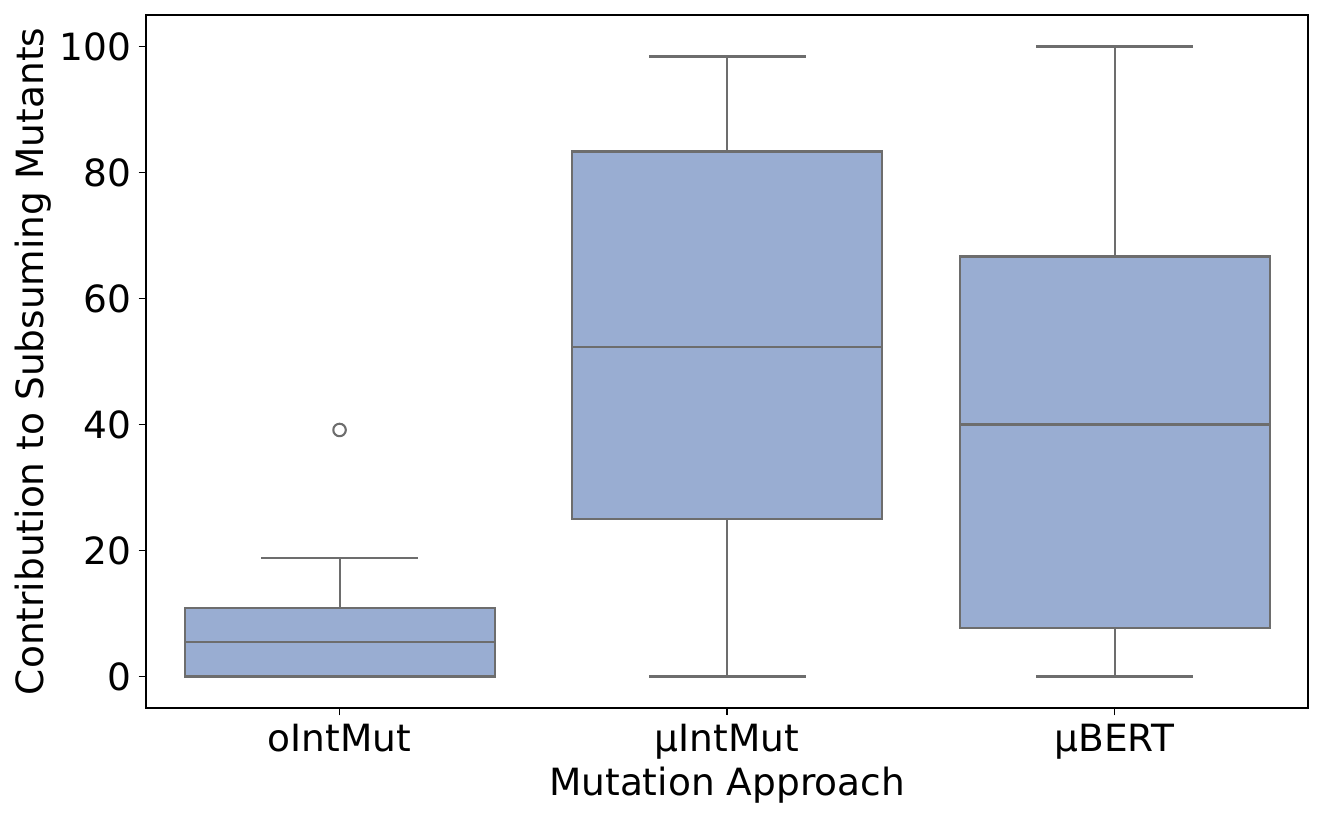}
\caption{Subsuming mutants among the different approaches}
\label{fig:subsumption}
\end{figure}

\noindent
\textbf{Subsumption:} To study the complementarity and subsumption between the different approaches, we merge their generated mutants and compute their subsuming set. 
Then we compute the percentage of subsuming mutants provided by each approach within this set.
We plot the obtained results for every task from our dataset in Figure~\ref{fig:subsumption}. 
The boxplots show that the three approaches contribute to forming the subsuming set with ratios varying between 0\% to 100\% for \toolname and \mbert, and between 0\% to 20\% for \llmo.
This indicates that for some tasks, one approach is subsuming the others while being totally subsumed for other tasks, however no approach subsumes always the other ones.

The boxplots depict also a large advantage for \toolname and \mbert over \llmo contributing on average by 53.3\%, 39.5\% and 7.2\% of the subsuming mutants respectively. This difference can be explained by the limited number of mutants generated by \llmo, which produced 4 and 9 times less mutants than \mbert and \toolname, as indicated in Table~\ref{tab:killed_mutants}.


\subsection{RQ3: How effective is intent-based mutation testing in comparison to syntactic-based one?}

\begin{figure*}[!h]
\centering
\includegraphics[width=0.95\textwidth]{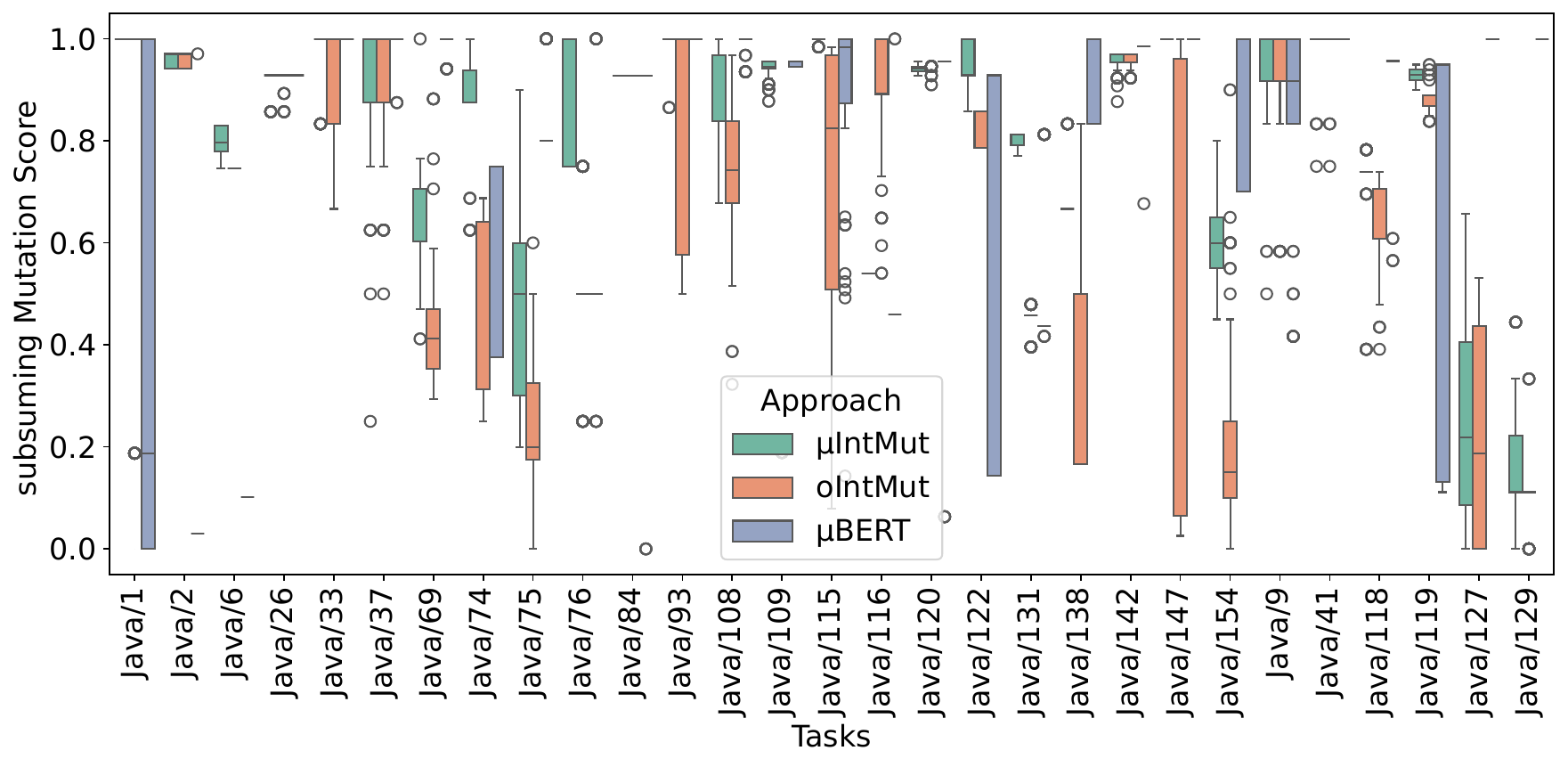}
\caption{\centering Distribution of subsuming mutation scores, on the reference set of subsuming mutants,  achieved by test suites selected based on each of the three approaches.}
\label{fig:ocs}
\end{figure*}

To have a common base of comparison between the approaches, we merge all mutants in one set and keep only the subsuming ones. Then we measure the subsuming mutation scores achieved by test suites designed to kill all the mutations of one approach at a time. Figure \ref{fig:ocs} presents the objective comparison scores (subsuming mutation score) scored by each of the three approaches.

The comparison was conducted 100 times, resulting in 100 distinct scores per case. 
We illustrate the distributions using boxplots, from which we observe a variance across problems in the percentage of mutants killed by the tests selected from \toolname mutants, with some tasks even reaching 100\%, and the majority exhibiting higher percentages, unlike \llmo and \mbert where we see higher variance and lower values.
The score of \toolname is 0.817 indicating that tests selected based on \toolname kill 81.7\% of the subsuming mutants produced by the three approaches (\toolname, \mbert and \llmo together), compared to 0.77 and 0.66 scored by \mbert and \llmo respectively. It is noted that since the reference set of subsuming mutants includes all the mutations produced by all the approaches it is expected that all approaches achieve high scores. The important thing though is that none of the approaches subsumes the others and all are far from being adequate. For instance, \mbert misses 23\% of the faults that could be captured when considering all three approaches together. This indicates a large gap, especially by considering that \mbert is a strong approach, arguably as strong as (or stronger) traditional mutation testing \cite{khanfir2023efficient}. 


Table \ref{tab:subsuming-ms} records the number of tasks where the scored subsuming mutation score by each approach is below a certain threshold, \toolname has the least number of cases below 50\% with only 2, and 17 tasks above 90\%. \mbert has the highest number above 90\% in 19 cases while \llmo has the lowest. These results indicate that a large variance in the effectiveness of all approaches, with  \toolname having the least low performing cases. \mbert on the other hand has the most cases with the highest scores.   

\begin{table}[!h]
\centering
  \caption{\centering Number of case with reference subsuming mutation score under various thresholds}
  \label{tab:subsuming-ms}
  \begin{tabular}{p{2cm}|c|c|c|c|c}
    \toprule
    \textbf{Approach} & 50\% & 60\% & 70\% & 80\% & 90\% \\
    \midrule
    \toolname & 2 & 4 & 6 & 10 & 12 \\
    \llmo & 9 & 11 & 12 & 14 & 17 \\
    \mbert & 8 & 9 & 9 & 9 & 10 \\
    \bottomrule
  \end{tabular}
\end{table}

\section{Discussion}
Table \ref{tab:examples} shows examples of mutants produced by \toolname. The problem is Java/37 from Humaneval-x. 
The replaced token and its position within the intent are in blue, green represents the original part, and red denotes the mutated one. For example, in Mutated Intent 1, the word \textbf{even} is replaced with \textbf{odd} and in Mutated Intent 2, a digit is altered from 6 to 2.

We notice that in mutant 1 the solution begins by adding the values at even indices to the result list, and the values at odd indices in another list, it sorts this list, and then adds its elements to the result list at odd indices. This behavior is the exact opposite of the original intent, which retains the values of odd indices and sorts the ones of even indices. 


\begin{table}[ht!]
\caption{\centering Mutated intents and related mutations.}
\label{tab:examples}
\begin{tabular}{p{0.45\textwidth}}
\toprule
\texttt{This function takes a list l and returns} \\
\texttt{a list l' such that l' is identical} \\
\texttt{to l in the odd indices, while its values } \\
\texttt{at the even indices are equal to the values} \\
\texttt{ of the even indices of l, but sorted.} \\
\\
\texttt{>>> sortEven(Arrays.asList(1, 2, 3))} \\
\texttt{[1, 2, 3]} \\
\texttt{>>> sortEven(Arrays.asList(5, 6, 3, 4))} \\
\texttt{[3, 6, 5, 4]} \\
\midrule
Resulting Implementation:  \\
\midrule
\begin{minipage}{0.45\textwidth}
\begin{lstlisting}
List<Integer> result = new ArrayList<>(l);
List<Integer> evenIndices = new ArrayList<>();
for (int i = 0; i < l.size(); i++) {
    if (i % 2 == 0) {
        evenIndices.add(l.get(i));
    }
}Collections.sort(evenIndices);
for (int i = 0; i < l.size(); i += 2) {
    result.set(i, evenIndices.remove(0));
}return result;
\end{lstlisting}
\end{minipage}
\\
\midrule
\textbf{Mutated Intent 1}  \\
\midrule
\begin{minipage}{0.45\textwidth}
\begin{lstlisting}[language=diff]
@@ even : 135 - 138 @@ 
+ values at the even indices are equal to the 
- values at the odd indices are equal to the 
\end{lstlisting}
\end{minipage}
\\
\midrule
Resulting Mutant 1:  \\
\midrule
\begin{minipage}{0.45\textwidth}
\begin{lstlisting}
List<Integer> result = new ArrayList<>();
List<Integer> evenValues = new ArrayList<>();
for (int i = 0; i < l.size(); i++) {
    if (i % 2 == 0) { result.add(l.get(i)); } 
    else { evenValues.add(l.get(i));}
}Collections.sort(evenValues);
int evenIndex = 0;
for (int i = 0; i < l.size(); i++) {
    if (i % 2 != 0) {
        result.add(evenValues.get(evenIndex));
        evenIndex++;
    }
}return result;
\end{lstlisting}
\end{minipage}
\\
\midrule
\textbf{Mutated Intent 2}  \\
\midrule
\begin{minipage}{0.45\textwidth}
\begin{lstlisting}[language=diff]
@@ 6 : 305 - 305 @@ 
+     >>> sortEven(Arrays.asList(5, 6, 3, 4))
-     >>> sortEven(Arrays.asList(5, 2, 3, 4))
\end{lstlisting}
\end{minipage}
\\
\midrule
Resulting Mutant 2:  \\
\midrule
\begin{minipage}{0.45\textwidth}
\begin{lstlisting}
List<Integer> result = new ArrayList<>(l);
List<Integer> evenValues = new ArrayList<>();
for (int i = 0; i < l.size(); i++) {
    if (i % 2 == 0) 
        evenValues.add(l.get(i));
}Collections.sort(evenValues);
int evenIndex = 0;
for (int i = 0; i < l.size(); i++) {
    if (i % 2 == 1) {
        result.set(i, evenValues.get(evenIndex));
        evenIndex++;
    }
}return result;
\end{lstlisting}
\vspace{-0.65em}
\end{minipage}
\\
\bottomrule
\end{tabular}
\label{tab:Java-37}
\end{table}

\subsection{Threats to validity}
The generalization of our results forms a threat to the validity of our work. 
A first concern may be attributed to the ability of Language Models 
to generatlize to other unseen data. 
Meaning, that performance may be different on other tasks and programs. 
To mitigate those threats we have conducted our experimental study on an independent dataset, specifically built to mitigate this threat in evaluating the performance of Large Language Models on code related tasks. 
The dataset counts tupples of human written code, description and tests that have not been included in the LLMs training sets.
Nevertheless, we acknowledge that the obtained results may not generalise to other cases. 

Other threats may arise from the non-deterministic nature of LLMs. This threat does not concern BERT, however concerns our employed code generation LLM. For instance, GPT-3.5-turbo tends to produce different answers for the same question. Although, this may reduce the reproducibility of our study, in a sense where we may obtain other mutants, we do not expect it to have an impact on our general results. Particularely, as we generate multiple mutants for different programs, we believe that the overall outcomes of the study will remain unchanged.

Some threats may arise from our semantic comparison of mutants based on their failing tests. We assume that the tests provided by HumanEval+ are complete and exhaustive, allowing us to capture behavioural differences between different programs. Although this may not be always the case, we believe that these test suites are sufficiently strong for our study. Moreover, we use the same setup, running the same test cases for all mutants generated by the compared approaches, giving a same base of comparison for all approaches.



\section{Conclusion and future work}
We presented \toolname, a mutation testing approach that generates mutants based on the program's intent. We proposed two ways to achieve this, by mutating the intent, and by generating different implementations of the intents. Our results revealed that \toolname produces a set of complex and semantically diverse mutants, which are semantically unique when compared to \mbert, a syntax-based approach, reflecting our approach's ability to capture different types of faults than those generated by traditional approaches. 

\input{appendix_reduced}

\bibliographystyle{IEEEtran}
\balance
\bibliography{sample}
\end{document}

%% file: appendix_reduced.tex
\newpage
\onecolumn
\appendix[Subject Used]
\begin{longtable}{p{0.45\textwidth} p{0.45\textwidth}}
  \toprule
  \endfirsthead

  \midrule
  \endhead
  \midrule
  \endfoot
  \bottomrule
  \endlastfoot
  \scriptsize \textbf{Java/1} : 
    Input to this function is a string containing multiple groups of nested parentheses. Your goal is to separate those group into separate strings and return the list of those.
    Separate groups are balanced (each open brace is properly closed) and not nested within each other. 
    Ignore any spaces in the input string. \newline
    \hspace*{1em}\texttt{>>>} separateParenGroups(""( ) (( )) (( )( ))"")
    [""()"", ""(())"", ""(()())""]
 & 
  \scriptsize \textbf{Java/2}:
    Given a positive floating point number, it can be decomposed into an integer part (largest integer smaller than the given number) and decimals (leftover part always smaller than 1).
    \newline
    Return the decimal part of the number.
    \newline
    \hspace*{1em}\texttt{>>>} truncateNumber(3.5)  0.5
    \\
  \midrule
  \scriptsize \textbf{Java/6} 
    Input to this function is a string represented multiple groups for nested parentheses separated by spaces.For each of the group, output the deepest level of nesting of parentheses.
    
    E.g. (()()) has maximum two levels of nesting while ((())) has three.
    
    \hspace*{1em}\texttt{>>>} parseNestedParens(""(()()) ((())) () ((())()())"") [2, 3, 1, 3]
    &
  \scriptsize \textbf{Java/9}
    From a given list of integers, generate a list of rolling maximum element found until the given moment in the sequence.
    
    \hspace*{1em}\texttt{>>>} rollingMax(Arrays.asList(1, 2, 3, 2, 3, 4, 2))
    
    \hspace*{1em}[1, 2, 3, 3, 3, 4, 4]
    \\
  \midrule  
  \scriptsize \textbf{Java/26}
    From a list of integers, remove all elements that occur more than once.
    Keep order of elements left the same as in the input.
    \newline
    \hspace*{1em}\texttt{>>>} removeDuplicates(Array.asList(1, 2, 3, 2, 4))
    \newline
    \hspace*{1em}[1, 3, 4]
    &
  \scriptsize \textbf{Java/33}
    This function takes a list l and returns a list l' such that l' is identical to l in the indicies that are not divisible by three, while its values at the indicies that are divisible by three are equal to the values of the corresponding indicies of l, but sorted.
    \newline
    \hspace*{1em}\texttt{>>>} sortThird(Arrays.asList(1, 2, 3))
    {1em}[1, 2, 3]
    \newline
    \hspace*{1em}\texttt{>>>} sortThird(Arrays.asList(5, 6, 3, 4, 8, 9, 2))
   [2, 6, 3, 4, 8, 9, 5]
    \\
  \midrule
  \scriptsize \textbf{Java/37} 
    This function takes a list l and returns a list l' such that l' is identical to l in the odd indicies, while its values at the even indicies are equal to the values of the even indicies of l, but sorted.
    \newline
    \hspace*{1em}\texttt{>>>} sortEven(Arrays.asList(1, 2, 3))
    \newline
    \hspace*{1em}[1, 2, 3]
    \newline
    \hspace*{1em}\texttt{>>>} sortEven(Arrays.asList(5, 6, 3, 4))
    \newline
    \hspace*{1em}[3, 6, 5, 4]
    &
  \scriptsize \textbf{Java/41}
    Imagine a road that's a perfectly straight infinitely long line.
    n cars are driving left to right; simultaneously, a different set of n cars
    are driving right to left.   The two sets of cars start out being very far from each other. All cars move at the same speed. Two cars are said to collide
    when a car that's moving left to right hits a car that's moving right to left.
    However, the cars are infinitely sturdy and strong; as a result, they continue moving in their trajectory as if they did not collide.
    This function outputs the number of such collisions.
    \\
    \midrule
  \scriptsize \textbf{Java/69}
    You are given a non-empty list of positive integers. Return the greatest integer that is greater than
    zero, and has a frequency greater than or equal to the value of the integer itself.
    The frequency of an integer is the number of times it appears in the list.
    If no such a value exists, return -1.
    \newline
    \hspace*{1em}Examples: search(Arrays.asList(4, 1, 2, 2, 3, 1)) == 2
    \newline
    \hspace*{1em}search(Arrays.asList(1, 2, 2, 3, 3, 3, 4, 4, 4)) == 3
    \newline
    \hspace*{1em}search(Arrays.asList(5, 5, 4, 4, 4)) == -1
   &

  \scriptsize \textbf{Java/74} 
    Write a function that accepts two lists of strings and returns the list that has
    total number of chars in all strings of the list less than the other list.
    if the two lists have the same number of chars, return the first list.
    \newline
    \hspace*{1em}Examples:
    \newline
    totalMatch(Arrays.asList(), Arrays.asList()) -> []
    \newline
    totalMatch(Arrays.asList("hi", "admin"), Arrays.asList("hI", "Hi"))\texttt{->}["hI", "Hi"]
    \newline
    totalMatch(Arrays.asList("4"), Arrays.asList("1", "2", "3", "4", "5"))\texttt{->}["4"]
    \\
  \midrule
  \scriptsize \textbf{Java/75}
    Write a function that returns true if the given number is the multiplication of 3 prime numbers
    and false otherwise.
    Knowing that (a) is less than 100.
    \newline
    \hspace*{1em}Example:
    \newline
    \hspace*{1em}isMultiplyPrime(30) == true
    \newline
    \hspace*{1em}30 = 2 * 3 * 5
    & 
  \scriptsize \textbf{Java/76} 
    Your task is to write a function that returns true if a number x is a simple
    power of n and false in other cases.
    x is a simple power of n if n**int=x
    \newline
    \hspace*{1em}For example:
    \newline
    \hspace*{1em}isSimplePower(1, 4) \texttt{=>} true

    \hspace*{1em}isSimplePower(5, 3) \texttt{=>} false
    \\
  \midrule
  \scriptsize \textbf{Java/84}
    Given a positive integer N, return the total sum of its digits in binary.
    \newline
    \hspace*{1em}Example:
        \newline
        \hspace*{2em}For N = 1000, the sum of digits will be 1 the output should be ""1"".
        \newline
        \hspace*{2em}For N = 150, the sum of digits will be 6 the output should be ""110"".
        \newline
        \hspace*{2em}For N = 147, the sum of digits will be 12 the output should be ""1100"".
    \newline
    Variables: @N integer
    \newline
            \hspace*{2em}Constraints: 0 \texttt{<=} \texttt{<=} 10000.
    \newline
    Output:
a string of binary number
    &
    \scriptsize \textbf{Java/93}	    
    Write a function that takes a message, and encodes in such a
    way that it swaps case of all letters, replaces all vowels in
    the message with the letter that appears 2 places ahead of that
    vowel in the english alphabet.
    Assume only letters.

    Examples:

    \hspace*{1em}\texttt{>>>} encode(""test"")
    
    \hspace*{1em}""TGST""
    
    \hspace*{1em}\texttt{>>>} encode(""This is a message"")
    
    \hspace*{1em}""tHKS KS C MGSSCGG""
        \\
   \midrule
    \scriptsize \textbf{Java/108} 
    Write a function countNums which takes an array of integers and returns
    the number of elements which has a sum of digits > 0.
    If a number is negative, then its first signed digit will be negative:
    \newline
    \hspace*{1em} e.g. -123 has signed digits -1, 2, and 3.
    \newline
    \hspace*{1em}\texttt{>>>} countNums(Arrays.asList()) == 0
    \newline
    \hspace*{1em}\texttt{>>>} countNums(Arrays.asList(-1, 11, -11)) == 1
    \newline
    \hspace*{1em}\texttt{>>>} countNums(Arrays.asList(1, 1, 2)) == 3
    
    \scriptsize \textbf{Java/109} 
    We have an array 'arr' of N integers arr[1], arr[2], ..., arr[N].The
    numbers in the array will be randomly ordered. Your task is to determine if
    it is possible to get an array sorted in non-decreasing order by performing
    the following operation on the given array:
    \newline
        \hspace*{1em}You are allowed to perform right shift operation any number of times.
    \newline
    One right shift operation means shifting all elements of the array by one
    position in the right direction. The last element of the array will be moved to
    the starting position in the array i.e. 0th index.
    \newline
    If it is possible to obtain the sorted array by performing the above operation
    then return true else return False.
    If the given array is empty then return true.
    \newline
    \hspace*{1em}Note: The given list is guaranteed to have unique elements.
    \newline
    For Example:
    \newline
    \hspace*{1em}moveOneBall(Arrays.asList(3, 4, 5, 1, 2))\hspace*{1em}\texttt{==>}true
    \newline
    \hspace*{1em}Explanation: By performin 2 right shift operations, non-decreasing order can
                 be achieved for the given array.
    \newline
    \hspace*{1em}moveOneBall(Arrays.asList(3, 5, 4, 1, 2))\hspace*{1em}\texttt{==>}False
    \newline
    \hspace*{1em}Explanation:It is not possible to get non-decreasing order for the given
                array by performing any number of right shift operations."
    &    
    \scriptsize \textbf{Java/115} 
    You are given a rectangular grid of wells. Each row represents a single well,
    and each 1 in a row represents a single unit of water.
    Each well has a corresponding bucket that can be used to extract water from it,
    and all buckets have the same capacity.
    \newline
    Your task is to use the buckets to empty the wells.
    \newline
    Output the number of times you need to lower the buckets.
    \newline
    \hspace*{1em}Example 1:
    \newline
        \hspace*{2em}Input: \hspace*{2em} grid : [[0,0,1,0], [0,1,0,0], [1,1,1,1]], bucket\_capacity : 1
            \newline
        \hspace*{2em}Output: 6
    \newline
    \hspace*{1em}Example 2:
    \newline
        \hspace*{2em}Input: \hspace*{2em} grid : [[0,0,1,1], [0,0,0,0], [1,1,1,1], [0,1,1,1]], bucket\_capacity : 2
            \newline
        \hspace*{2em}Output: 5
\newline
    \hspace*{1em}Example 3:
    \newline
        \hspace*{2em}Input: \hspace*{2em}grid : [[0,0,0], [0,0,0]], bucket\_capacity : 5
            \newline
        \hspace*{2em}Output: 0
\newline
 Constraints:
    \newline
        \hspace*{1em}* all wells have the same length
        \hspace*{2em}* 1 \texttt{<=} grid.length \texttt{<=} $10^2$
        
        \hspace*{1em}* 1 \texttt{<=} grid[:,1].length \texttt{<=} $10^2$ \hspace*{2em}* grid[i][j] \texttt{->} 0 | 1
        \newline
        \hspace*{1em}* 1 \texttt{<=} capacity \texttt{<=} 10
            \\

    \scriptsize \textbf{Java/118} 
    You are given a word. Your task is to find the closest vowel that stands between
    two consonants from the right side of the word (case sensitive).
    Vowels in the beginning and ending doesn't count. Return empty string if you didn't
    find any vowel met the above condition.
    You may assume that the given string contains English letter only.
    
    \hspace*{1em}Example:
    
    \hspace*{1em}getClosestVowel(""yogurt"") \texttt{==>} ""u""
    
    \hspace*{1em}getClosestVowel(""FULL"") \texttt{==>} ""U""
    
    \hspace*{1em}getClosestVowel(""quick"") \texttt{==>} """"
    
    \hspace*{1em}getClosestVowel(""ab"") \texttt{==>} """"
    &

    \scriptsize \textbf{Java/119}	   
    You are given a list of two strings, both strings consist of open parentheses ""("" or close parentheses "")"" only.
    Your job is to check if it is possible to concatenate the two strings in
    some order, that the resulting string will be good.
    A string S is considered to be good if and only if all parentheses in S
    are balanced. For example: the string ""(())()"" is good, while the string
    ""())"" is not.
    Return ""Yes"" if there""s a way to make a good string, and return ""No"" otherwise.

    \hspace*{1em}Examples:
    
    \hspace*{1em}matchParens(Arrays.asList(""()("", "")"")) == ""Yes""
    
    \hspace*{1em}matchParens(Arrays.asList("")"", "")"")) == ""No""
        \\
    \midrule

    \scriptsize \textbf{Java/120}	 Given an array arr of integers and a positive integer k, return a sorted list
    of length k with the maximum k numbers in arr.

    \hspace*{1em}Example 1: Input:arr = [-3, -4, 5], k = 3
        
        \hspace*{2em}Output: [-4, -3, 5]

    \hspace*{1em}Example 2: Input: arr = [4, -4, 4], k = 2
        
        \hspace*{2em}Output: [4, 4]

    \hspace*{1em}Example 3: Input: arr = [-3, 2, 1, 2, -1, -2, 1], k = 1
    
        \hspace*{2em}Output: [2]

    Note: * 0 \texttt{<=} k \texttt{<=} len(arr)

        \hspace*{1em}* The length of the array will be in the range of [1, 1000].
    
        \hspace*{1em}* The elements in the array will be in the range of [-1000, 1000].    
    &
    
    \scriptsize \textbf{Java/122}
    Given a non-empty array of integers arr and an integer k, return
    the sum of the elements with at most two digits from the first k elements of arr.

    \hspace*{1em}Example:

        \hspace*{2em}Input: arr = [111,21,3,4000,5,6,7,8,9], k = 4
        
        \hspace*{2em}Output: 24 \# sum of 21 + 3

    Constraints:
    
        \hspace*{1em}1. 1 \texttt{<=} len(arr) \texttt{<=} 100
        
        \hspace*{1em}2. 1 \texttt{<=} k \texttt{<=} len(arr)
        \\
    \midrule

    \scriptsize \textbf{Java/127} 
    
    You are given two intervals,
    where each interval is a pair of integers. For example, interval = (start, end) = (1, 2).
    The given intervals are closed which means that the interval (start, end)
    includes both start and end.
    For each given interval, it is assumed that its start is less or equal its end.
    Your task is to determine whether the length of intersection of these two
    intervals is a prime number.
    
    Example, the intersection of the intervals (1, 3), (2, 4) is (2, 3)
    which its length is 1, which not a prime number.
    If the length of the intersection is a prime number, return ""YES"",
    otherwise, return ""NO"".
    If the two intervals don't intersect, return ""NO"".

    [input/output] samples:
    
    \hspace*{1em}intersection((1, 2), (2, 3)) \texttt{==>} ""NO""
    
    \hspace*{1em}intersection((-1, 1), (0, 4)) \texttt{==>} ""NO""
    
    \hspace*{1em}intersection((-3, -1), (-5, 5)) \texttt{==>} ""YES""
    &
    \scriptsize \textbf{Java/129}	
    Given a grid with N rows and N columns (N \texttt{>=} 2) and a positive integer k,
    each cell of the grid contains a value. Every integer in the range [1, N * N]
    inclusive appears exactly once on the cells of the grid.
    You have to find the minimum path of length k in the grid. You can start
    from any cell, and in each step you can move to any of the neighbor cells,
    in other words, you can go to cells which share an edge with you current
    cell.
    Please note that a path of length k means visiting exactly k cells (not
    necessarily distinct).
    You CANNOT go off the grid.
    A path A (of length k) is considered less than a path B (of length k) if
    after making the ordered lists of the values on the cells that A and B go
    through (let's call them lst\_A and lst\_B), lst\_A is lexicographically less
    than lst\_B, in other words, there exist an integer index i (1 \texttt{<=} i \texttt{<=} k)
    such that lst\_A[i] \texttt{<} lst\_B[i] and for any j (1 \texttt{<=} j \texttt{<} i) we have
    lst\_A[j] = lst\_B[j].
    It is guaranteed that the answer is unique.
    Return an ordered list of the values on the cells that the minimum path go through.

    Examples:

        \hspace*{1em}Input: grid = [ [1,2,3], [4,5,6], [7,8,9]], k = 3
        
        \hspace*{1em}Output: [1, 2, 1]

        \hspace*{1em}Input: grid = [ [5,9,3], [4,1,6], [7,8,2]], k = 1
        
        \hspace*{1em}Output: [1]
    \\
    \midrule

    \scriptsize \textbf{Java/131}	
    Given a positive integer n, return the product of the odd digits.
    Return 0 if all digits are even.
    
    For example:
    
    \hspace*{1em}digits(1)  == 1
    \hspace*{1em}
    \hspace*{1em}digits(4)  == 0
    \hspace*{1em}
    \hspace*{1em}digits(235) == 15
    & 

    \scriptsize \textbf{Java/138}
    Evaluate whether the given number n can be written as the sum of exactly 4 positive even numbers
    
    Example : \hspace*{1em}isEqualToSumEven(4) == false
    \hspace*{1em}
    \hspace*{1em}isEqualToSumEven(6) == false
    \hspace*{1em}
    \hspace*{1em}isEqualToSumEven(8) == true
    \\
    \midrule
    
    \scriptsize \textbf{Java/142}	
    This function will take a list of integers. For all entries in the list, the function shall square the integer entry if its index is a
    multiple of 3 and will cube the integer entry if its index is a multiple of 4 and not a multiple of 3. The function will not
    change the entries in the list whose indexes are not a multiple of 3 or 4. The function shall then return the sum of all entries.

    Examples:
    
    \hspace*{1em}For lst = [1,2,3] the output should be 6
    
    \hspace*{1em}For lst = []  the output should be 0
    
    \hspace*{1em}For lst = [-1,-5,2,-1,-5]  the output should be -126
    
    &
    \scriptsize \textbf{Java/147}	
    You are given positive integer n. You have to create  integer array a of length n.
    
        \hspace*{1em}For each i (1 \texttt{<=} i \texttt{<=} n), the value of a[i] = i * i - i + 1.
        
        \hspace*{1em}Return the number of triples (a[i], a[j], a[k]) of a where i \texttt{<} j \texttt{<} k,
        
    and a[i] + a[j] + a[k] is a multiple of 3.

    Example :
    
        \hspace*{1em}Input: n = 5\hspace*{1em}Output: 1
    
        \hspace*{1em}Explanation: \hspace*{1em}a = [1, 3, 7, 13, 21]
    
        \hspace*{1em}The only valid triple is (1, 7, 13).
        \\
    \midrule

    \multicolumn{2}{p{17cm}}{\raggedright
    \scriptsize \textbf{Java/154}	
    You are given 2 words. You need to return true if the second word or any of its rotations is a substring in the first word
    
    \hspace*{1em}cycpatternCheck(""abcd"",""abd"") \texttt{=>} false
    
    \hspace*{1em}cycpatternCheck(""hello"",""ell"") \texttt{=>} true
    
    \hspace*{1em}cycpatternCheck(""whassup"",""psus"") \texttt{=>} false
    
    \hspace*{1em}cycpatternCheck(""abab"",""baa"") \texttt{=>} true
    
    \hspace*{1em}cycpatternCheck(""efef"",""eeff"") \texttt{=>} false
    
    \hspace*{1em}cycpatternCheck(""himenss"",""simen"") \texttt{=>} true
}
\end{longtable}

\twocolumn

%% file: sample.bib
@inproceedings{zheng2023codegeex,
  title={Codegeex: A pre-trained model for code generation with multilingual benchmarking on humaneval-x},
  author={Zheng, Qinkai and Xia, Xiao and Zou, Xu and Dong, Yuxiao and Wang, Shan and Xue, Yufei and Shen, Lei and Wang, Zihan and Wang, Andi and Li, Yang and others},
  booktitle={Proceedings of the 29th ACM SIGKDD Conference on Knowledge Discovery and Data Mining},
  pages={5673--5684},
  year={2023}
}

@article{chen2021evaluating,
  title={Evaluating large language models trained on code},
  author={Chen, Mark and Tworek, Jerry and Jun, Heewoo and Yuan, Qiming and Pinto, Henrique Ponde De Oliveira and Kaplan, Jared and Edwards, Harri and Burda, Yuri and Joseph, Nicholas and Brockman, Greg and others},
  journal={arXiv:2107.03374},
  year={2021}
}

@article{liu2024your,
  title={Is your code generated by chatgpt really correct? rigorous evaluation of large language models for code generation},
  author={Liu, Jiawei and Xia, Chunqiu Steven and Wang, Yuyao and Zhang, Lingming},
  journal={Advances in Neural Information Processing Systems},
  volume={36},
  year={2024}
}

@inproceedings{DBLP:conf/icst/DegiovanniP22,
  author    = {Renzo Degiovanni and
               Mike Papadakis},
  title     = {{\(\mathrm{\mu}\)}Bert: Mutation Testing using Pre-Trained Language
               Models},
  booktitle = {15th {IEEE} International Conference on Software Testing, Verification
               and Validation Workshops {ICST} Workshops},
  pages     = {160--169},
  __publisher = {{IEEE}},
  year      = {2022},
  __url       = {https://doi.org/10.1109/ICSTW55395.2022.00039},
  __doi       = {10.1109/ICSTW55395.2022.00039},
}

@inproceedings{tian2022learning,
  title={Learning to construct better mutation faults},
  author={Tian, Zhao and Chen, Junjie and Zhu, Qihao and Yang, Junjie and Zhang, Lingming},
  booktitle={Proceedings of the International Conference on Automated Software Engineering},
  pages={1--13},
  year={2022}
}

@article{devlin2018bert,
  title={BERT: Pre-training of Deep Bidirectional Transformers for Language Understanding},
  author={Devlin, Jacob and Chang, Ming-Wei and Lee, Kenton and Toutanova, Kristina},
  journal={arXiv:1810.04805},
  year={2018}
}

@article{khanfir2020ibir,
author = {Khanfir, Ahmed and Koyuncu, Anil and Papadakis, Mike and Cordy, Maxime and Bissyand\'{e}, Tegawende F. and Klein, Jacques and Le Traon, Yves},
title = {IBiR: Bug Report Driven Fault Injection},
year = {2022},
publisher = {Association for Computing Machinery},
address = {New York, NY, USA},
issn = {1049-331X},
__url = {https://doi.org/10.1145/3542946},
__doi = {10.1145/3542946},
journal = {ACM Trans. Softw. Eng. Methodol.},
month = {may},
keywords = {Mutation, Information Retrieval, Bug Reports, Fault Injection}
}

@book{0020331,
  author    = {Paul Ammann and
               Jeff Offutt},
  title     = {Introduction to Software Testing},
  publisher = {Cambridge University Press},
  year      = {2008},
  __url       = {https://doi.org/10.1017/CBO9780511809163},
  __doi       = {10.1017/CBO9780511809163},
  isbn      = {978-0-521-88038-1},
}

@inproceedings{KurtzADOD2014,
author = {Kurtz, Bob and Ammann, Paul and Delamaro, Marcio E. and Offutt, Jeff and Deng, Lin},
title = {Mutant Subsumption Graphs},
year = {2014},
__isbn = {9781479957903},
__publisher = {IEEE Computer Society},
__address = {USA},
__url = {https://doi.org/10.1109/ICSTW.2014.20},
__doi = {10.1109/ICSTW.2014.20},
abstract = {Mutation testing researchers have long known that many generated mutants are not needed.
This paper develops a graph model to describe redundancy among mutations. We define
"true" subsumption, a relation that practicing test engineers would like to have,
but cannot due to issues of computability. We also define dynamic subsumption and
static subsumption as approximations of "true" subsumption. We explore the properties
of the approximate subsumption relations in the context of a small example. We suggest
possible uses for subsumption graphs.},
booktitle = {International Conference on Software Testing, Verification, and Validation Workshops {ICSTW}},
pages = {176–185},
_numpages = {10},
_keywords = {mutation testing, subsumption},
__series = {ICSTW '14}
}

@inproceedings{PapadakisJHT15,
  author       = {Mike Papadakis and
                  Yue Jia and
                  Mark Harman and
                  Yves Le Traon},
  __editor       = {Antonia Bertolino and
                  Gerardo Canfora and
                  Sebastian G. Elbaum},
  title        = {Trivial Compiler Equivalence: {A} Large Scale Empirical Study of a
                  Simple, Fast and Effective Equivalent Mutant Detection Technique},
  booktitle    = {37th {IEEE/ACM} International Conference on Software Engineering,
                  {ICSE}},
  pages        = {936--946},
  __publisher    = {{IEEE} Computer Society},
  year         = {2015},
  __url          = {https://doi.org/10.1109/ICSE.2015.103},
  __doi          = {10.1109/ICSE.2015.103},
}

@inproceedings{DBLP:conf/emnlp/FengGTDFGS0LJZ20,
  author    = {Zhangyin Feng and
               Daya Guo and
               Duyu Tang and
               Nan Duan and
               Xiaocheng Feng and
               Ming Gong and
               Linjun Shou and
               Bing Qin and
               Ting Liu and
               Daxin Jiang and
               Ming Zhou},
  title     = {CodeBERT: {A} Pre-Trained Model for Programming and Natural Languages},
  booktitle = {Conference on Empirical Methods in Natural
               Language Processing: Findings, {EMNLP}},
  __series    = {Findings of {ACL}},
  __volume    = {{EMNLP}},
  pages     = {1536--1547},
  __publisher = {Association for Computational Linguistics},
  year      = {2020},
  __url       = {https://doi.org/10.18653/v1/2020.findings-emnlp.139},
  __doi       = {10.18653/v1/2020.findings-emnlp.139},
}

@article{KintisPM15,
  author    = {Marinos Kintis and
               Mike Papadakis and
               Nicos Malevris},
  title     = {Employing second-order mutation for isolating first-order equivalent
               mutants},
  journal   = {Softw. Test. Verification Reliab.},
  volume    = {25},
  number    = {5-7},
  pages     = {508--535},
  year      = {2015},
  __url       = {https://doi.org/10.1002/stvr.1529},
  __doi       = {10.1002/stvr.1529},
}

@inproceedings{MarcozziBKPPC18,
  author       = {Micha{\"{e}}l Marcozzi and
                  S{\'{e}}bastien Bardin and
                  Nikolai Kosmatov and
                  Mike Papadakis and
                  Virgile Prevosto and
                  Lo{\"{\i}}c Correnson},
  __editor       = {Michel Chaudron and
                  Ivica Crnkovic and
                  Marsha Chechik and
                  Mark Harman},
  title        = {Time to clean your test objectives},
  booktitle    = {International Conference on Software Engineering,
                  {ICSE}},
  pages        = {456--467},
  __publisher    = {{ACM}},
  year         = {2018},
  __url          = {https://doi.org/10.1145/3180155.3180191},
  __doi          = {10.1145/3180155.3180191},
}

@inproceedings{ChekamPTH17,
  author    = {Thierry Titcheu Chekam and
               Mike Papadakis and
               Yves Le Traon and
               Mark Harman},
  title     = {An empirical study on mutation, statement and branch coverage fault
               revelation that avoids the unreliable clean program assumption},
  booktitle = { International Conference on Software Engineering,
               {ICSE}},
  pages     = {597--608},
  year      = {2017},
  __url       = {https://doi.org/10.1109/ICSE.2017.61},
  __doi       = {10.1109/ICSE.2017.61},
}

@article{PapadakisK00TH19,
  author    = {Mike Papadakis and
               Marinos Kintis and
               Jie Zhang and
               Yue Jia and
               Yves Le Traon and
               Mark Harman},
  title     = {Chapter Six - Mutation Testing Advances: An Analysis and Survey},
  journal   = {Advances in Computers},
  volume    = {112},
  pages     = {275--378},
  year      = {2019},
  __url       = {https://doi.org/10.1016/bs.adcom.2018.03.015},
  __doi       = {10.1016/bs.adcom.2018.03.015},
}

@article{ChekamPBTS20,
  author    = {Thierry Titcheu Chekam and
               Mike Papadakis and
               Tegawend{\'{e}} F. Bissyand{\'{e}} and
               Yves Le Traon and
               Koushik Sen},
  title     = {Selecting fault revealing mutants},
  journal   = {Empirical Software Engineering},
  volume    = {25},
  number    = {1},
  pages     = {434--487},
  year      = {2020},
  __url       = {https://doi.org/10.1007/s10664-019-09778-7},
  __doi       = {10.1007/s10664-019-09778-7},
}

@inproceedings{pitest,
author = {Coles, Henry and Laurent, Thomas and Henard, Christopher and Papadakis, Mike and Ventresque, Anthony},
title = {PIT: A Practical Mutation Testing Tool for Java (Demo)},
year = {2016},
__isbn = {9781450343909},
__publisher = {Association for Computing Machinery},
__address = {New York, NY, USA},
__url = {https://doi.org/10.1145/2931037.2948707},
__doi = {10.1145/2931037.2948707},
__abstract = { Mutation testing introduces artificial defects to measure the adequacy of testing. In case candidate tests can distinguish the behaviour of mutants from that of the original program, they are considered of good quality -- otherwise developers need to design new tests. While, this method has been shown to be effective, industry-scale code challenges its applicability due to the sheer number of mutants and test executions it requires. In this paper we present PIT, a practical mutation testing tool for Java, applicable on real-world codebases. PIT is fast since it operates on bytecode and optimises mutant executions. It is also robust and well integrated with development tools, as it can be invoked through a command line interface, Ant or Maven. PIT is also open source and hence, publicly available at url{http://pitest.org/} },
booktitle = {Proceedings of the 25th International Symposium on Software Testing and Analysis},
pages = {449–452},
numpages = {4},
}

@inproceedings{PapadakisCT18,
  author    = {Mike Papadakis and
               Thierry Titcheu Chekam and
               Yves Le Traon},
  title     = {Mutant Quality Indicators},
  booktitle = {2018 {IEEE} International Conference on Software Testing, Verification
               and Validation Workshops},
  pages     = {32--39},
  __publisher = {{IEEE} Computer Society},
  year      = {2018},
  __url       = {http://doi.ieeecomputersociety.org/10.1109/ICSTW.2018.00025},
  __doi       = {10.1109/ICSTW.2018.00025},
}

@inproceedings{SemSeed,
author = {Patra, Jibesh and Pradel, Michael},
title = {Semantic Bug Seeding: A Learning-Based Approach for Creating Realistic Bugs},
year = {2021},
booktitle    = {{ESEC/FSE} Joint European Software Engineering Conference
                  and Symposium on the Foundations of Software Engineering},
__isbn = {9781450385626},
__publisher = {Association for Computing Machinery},
__address = {New York, NY, USA},
__url = {https://doi.org/10.1145/3468264.3468623},
__doi = {10.1145/3468264.3468623},
pages = {906–918},
__numpages = {13},
__keywords = {bug injection, dataset, machine learning, token embeddings, bugs},
__location = {Athens, Greece},
__series = {ESEC/FSE 2021}
}

@inproceedings{DeepMutation,
author = {Tufano, Michele and Kimko, Jason and Wang, Shiya and Watson, Cody and Bavota, Gabriele and Di Penta, Massimiliano and Poshyvanyk, Denys},
title = {DeepMutation: A Neural Mutation Tool},
year = {2020},
__isbn = {9781450371223},
__publisher = {Association for Computing Machinery},
__address = {New York, NY, USA},
__url = {https://doi.org/10.1145/3377812.3382146},
__doi = {10.1145/3377812.3382146},
booktitle = {International Conference on Software Engineering: Companion Proceedings},
pages = {29–32},
__numpages = {4},
__keywords = {neural networks, mutation testing, software testing},
__location = {Seoul, South Korea},
series = {ICSE}
}

@article{mujava,
  author    = {Yu{-}Seung Ma and
               Jeff Offutt and
               Yong Rae Kwon},
  title     = {MuJava: an automated class mutation system},
  journal   = {Softw. Test. Verification Reliab.},
  volume    = {15},
  number    = {2},
  pages     = {97--133},
  year      = {2005},
  __url       = {https://doi.org/10.1002/stvr.308},
  __doi       = {10.1002/stvr.308},
}

@inproceedings{major,
author={T. Laurent and M. Papadakis and M. Kintis and C. Henard and Y. L. Traon and A. Ventresque},
booktitle={2017 IEEE International Conference on Software Testing, Verification and Validation (ICST)},
title={Assessing and Improving the Mutation Testing Practice of PIT},
year={2017},
volume={},
number={},
pages={430-435},
keywords={program testing;software tools;PIT;community standards;major;muJava;mutation testing practice;mutation testing tools;mutation-adequate test cases;mutation-based experiments;software testing;Benchmark testing;Java;Robustness;Software testing;Syntactics;Tools},
__doi={10.1109/ICST.2017.47},
ISSN={},
month={March},}

@article{KintisPPVMT18,
  author    = {Marinos Kintis and
               Mike Papadakis and
               Andreas Papadopoulos and
               Evangelos Valvis and
               Nicos Malevris and
               Yves Le Traon},
  title     = {How effective are mutation testing tools? An empirical analysis of
               Java mutation testing tools with manual analysis and real faults},
  journal   = {Empir. Softw. Eng.},
  volume    = {23},
  number    = {4},
  pages     = {2426--2463},
  year      = {2018},
  __url       = {https://doi.org/10.1007/s10664-017-9582-5},
  __doi       = {10.1007/s10664-017-9582-5},
}

@misc{copilot,
    title = {GitHub Copilot},
    howpublished = {\url{https://github.com/features/copilot}},
}

@misc{codewhisperer,
    title = {Amazon CodeWhisperer},
    howpublished = {\url{https://aws.amazon.com/codewhisperer/}},
}

@article{openAIcodex,
  title={Evaluating large language models trained on code.(2021)},
  author={Chen, Mark and Tworek, Jerry and Jun, Heewoo and Yuan, Qiming and de Oliveira Pinto, Henrique Ponde and Kaplan, J and Edwards, H and Burda, Y and Joseph, N and Brockman, G and others},
  journal={arXiv:2107.03374},
  year={2021}
}

@article{OffuttLRUZ96,
  author    = {A. Jefferson Offutt and
               Ammei Lee and
               Gregg Rothermel and
               Roland H. Untch and
               Christian Zapf},
  title     = {An Experimental Determination of Sufficient Mutant Operators},
  journal   = {{ACM} Trans. Softw. Eng. Methodol.},
  volume    = {5},
  number    = {2},
  pages     = {99--118},
  year      = {1996},
  __url       = {https://doi.org/10.1145/227607.227610},
  __doi       = {10.1145/227607.227610},
}

@article{ojdanic2023syntactic,
  title={Syntactic versus semantic similarity of artificial and real faults in mutation testing studies},
  author={Ojdanic, Milos and Garg, Aayush and Khanfir, Ahmed and Degiovanni, Renzo and Papadakis, Mike and Le Traon, Yves},
  journal={IEEE Transactions on Software Engineering},
  volume={49},
  number={7},
  pages={3922--3938},
  year={2023},
  publisher={IEEE}
}

@inproceedings{ojdanic2023comparing,
  title={On comparing mutation testing tools through learning-based mutant selection},
  author={Ojdanic, Milos and Khanfir, Ahmed and Garg, Aayush and Degiovanni, Renzo and Papadakis, Mike and Le Traon, Yves},
  booktitle={2023 IEEE/ACM International Conference on Automation of Software Test (AST)},
  pages={35--46},
  year={2023},
  organization={IEEE}
}

@article{herzig2011untangling,
  title={Untangling changes},
  author={Herzig, Kim and Zeller, Andreas},
  journal={Unpublished manuscript, September},
  volume={37},
  pages={38--40},
  year={2011}
}

@inproceedings{tcap,
author = {Kaufman, Samuel J. and Featherman, Ryan and Alvin, Justin and Kurtz, Bob and Ammann, Paul and Just, Ren\'{e}},
title = {Prioritizing Mutants to Guide Mutation Testing},
year = {2022},
__isbn = {9781450392211},
__publisher = {Association for Computing Machinery},
__address = {New York, NY, USA},
__url = {https://doi-org.proxy.bnl.lu/10.1145/3510003.3510187},
__doi = {10.1145/3510003.3510187},
booktitle = {International Conference on Software Engineering},
pages = {1743–1754},
__numpages = {12},
__keywords = {mutant selection, test completeness advancement probability, TCAP, mutation testing, mutant utility, machine learning},
__location = {Pittsburgh, Pennsylvania},
__series = {ICSE '22}
}

@article{khanfir2023efficient,
  title={Efficient mutation testing via pre-trained language models},
  author={Khanfir, Ahmed and Degiovanni, Renzo and Papadakis, Mike and Traon, Yves Le},
  journal={arXiv:2301.03543},
  year={2023}
}
